\documentclass[reprint, secnumarabic, amssymb, amsmath, superscriptaddress, aps, prl, longbibliography]{revtex4-2}
\usepackage{graphicx}
\usepackage[colorlinks=true,urlcolor=blue]{hyperref}
\usepackage{hyperref}
\usepackage{xcolor}
\usepackage{physics}
\usepackage{amsmath}
\usepackage[utf8]{inputenc}
\hypersetup{
colorlinks,
linkcolor={red!50!black},
citecolor={blue!50!black},
urlcolor={blue!80!black}
}
\usepackage[normalem]{ulem}
\usepackage{rotating}
\usepackage{units}
\usepackage{subcaption}  
\usepackage{wrapfig} 
\DeclareUnicodeCharacter{2212}{-}

\begin{document}

\title{Collective charge excitations studied by electron energy-loss spectroscopy}

\author{Peter Abbamonte}
\email{abbamont@illinois.edu}
\affiliation{Department of Physics, University of Illinois at Urbana–Champaign, Urbana, IL 61801, USA}
\affiliation{Materials Research Laboratory, University of Illinois, Urbana, IL 61801, USA}

\author{J\"org Fink}
\email{J.Fink@ifw-dresden.de}
\affiliation{Leibniz Institute for Solid State and Materials Research  Dresden, Helmholtzstr. 20, D-01069 Dresden, Germany}
\affiliation {Institut f\"ur Festk\"orperphysik,  Technische Universit\"at Dresden, D-01062 Dresden, Germany}

\date{\today}

\begin{abstract}
The dynamic charge susceptibility, $\chi(q,\omega)$, is a fundamental observable of all materials, in one, two, and three dimensions, quantifying the collective charge modes, the ability of a material to screen charge, as well as its electronic compressibility. Here, we review the current state of efforts to measure this quantity using inelastic electron scattering, which historically has been called electron energy-loss spectroscopy (EELS). We focus on comparison between transmission (T-EELS) and reflection (R-EELS) geometries as applied to a selection of 3D conductors. While a great deal is understood about simple metals, measurements of more strongly interacting and strange metals are currently contradictory, with different groups obtaining fundamentally contradictory results, emphasizing the importance of improved EELS measurements. Further, current opportunities for improvement in EELS techniques are vast, with the most promising future development being in hemispherical and time-of-flight analyzers, as well as STEM instruments configured for high momentum resolution. We conclude that, despite more than half a century of work, EELS techniques are currently still in their infancy. 

\end{abstract}

\maketitle

\section{I. Introduction}\label{sec:intro}

Interacting electron systems, often referred to today as ``quantum materials", lie at the forefront of condensed matter physics, and have been the basis for a stunning array of discoveries over the last 30 years. Some highlights include Fe-based high temperature superconductivity~\cite{Kamihara2008,Johnston2010,Paglione2010}, 
topological insulators and semimetals
~\cite{Hasan2010,Lv2021}, 
a wide variety of broken symmetry phases including charge and spin density waves 
~\cite{Chen2016}, 
exciton condensates 
~\cite{Eisenstein2014,Kogar2017}, 
emergent fractionalized phases ~\cite{Eisenstein2014,Xu2023}, 
and superconductors with a variety of nontrivial order parameters ~\cite{Stewart2017,Madhavan2020}, 
in both 3D and 2D materials. Progress continues to accelerate, and many more discoveries are surely in store. 

The foundation of our understanding of quantum materials are advanced spectroscopies that measure the elementary excitations at the meV energy scale. Some of the most widely used are infrared spectroscopy ~\cite{Dressel2002,Basov2005}, 
angle-resolved photoemission spectroscopy (ARPES) ~\cite{Damascelli2003}, 
scanning tunneling microscopy and spectroscopy (STM or STS) ~\cite{Fischer2007},
inelastic neutron scattering (INS)~\cite{Balcar1989}, 
and non-resonant and resonant inelastic x-ray scattering 
(IXS~\cite{Schuelke2007} and RIXS~\cite{Ament2011}, respectively). 
All of these techniques detect excitations by measuring elementary Fermion or Boson response functions as a function of frequency as well as momentum or position, and are the basis for microscopic understanding of most materials. 

There is recent, growing interest in 
techniques for measuring the charge collective modes at the meV scale by inelastic scattering of electrons. 
In these methods, the electric field of an electron, moving in or near the surface of a material, probes the dynamic charge susceptibility by perturbing the density in the material. This method is usually referred to as electron energy-loss spectroscopy (EELS) or, sometimes, inelastic electron scattering (IES). Essentially, EELS is an extension of the Franck-Hertz experiment originally used to detect electronic excitations of gases
~\cite{Franck1913}.  
Energy losses in solids were demonstrated both for electrons reflected from surfaces
~\cite{Rudberg1936} and in transmission through foils~\cite{Ruthemann1941}. 
These losses  were later explained in terms of plasmons~\cite{Pines1952}, 
though any type of longitudinal charge collective mode can be detected using EELS techniques. 

The beauty of EELS is that it provides information not only on the energy but also on the momentum/wavelength dependence of excitations~\cite{Watanabe1956}. 
The spanned momentum range is of the order of the size of the Brillouin zone (BZ) of a typical material ($\approx 1$  \AA $^{-1}$), and the momentum resolution can be comparable to other wave vector resolved techniques like ARPES and INS. EELS should therefore be considered a standard technique for studying excitations in materials. 

EELS  in  transmission (T-EELS) was originally performed using dedicated spectrometers based on dispersive dipoles ~\cite{Raether1965,Daniels1970,Schnatterly1979,Schattschneider1950,Fink1989,Baltz1997,Roth2014}
or as an attachment on a scanning transmission electron microscope (STEM)
~\cite{Egerton1996,Rose2008,mkhoyan2007}.
The energy energy resolution in these instruments was in most cases not below $\Delta E \approx 0.1 $ eV, though, in one instance, resolution better than 20 meV was achieved using Wien filters
~\cite{Boersch1964,Schroeder1972}. 
EELS in reflection geometry, originally developed for surface science applications, is often referred to as HR-EELS (for ``high resolution" EELS) since it can achieve resolution $\Delta E \lesssim 10 $ meV~\cite{Ibach1982,Ibach1991,Ibach1993,Ibach2017,Ibach2013}. In this article, we will refer to such reflection measurements as ``R-EELS" to emphasize their difference from T-EELS. 

Over the last several decades, significant progress has been made in both approaches. Using aberration correction techniques in modern STEM instruments, resolution better than 10 meV has now been achieved in T-EELS at beam energies above 20 keV
~\cite{Krivanek2019,Krivanek2014}.
Further, R-EELS spectrometers have been constructed that achieve energy resolution $\Delta E \lesssim 1$ meV using toroidal lenses
~\cite{Ibach1993}. 
The latter instruments have now been configured 
to achieve full momentum tunability, comparable to ARPES or INS techniques, either by using eucentric sample goniometers
~\cite{Vig2017,Mitrano2018,Husain2019} 
or parallel readout using ARPES-type hemispherical analyzers \cite{Zhu2015,Ibach2017}. 
R-EELS techniques have recently detected a Bose condensation of excitons in TiSe$_2$ 
\cite{Kogar2017}, 
topological phonons in graphene
~\cite{Li2023},
and an acoustic plasmon or ``demon" excitation in Sr$_2$RuO$_4$ 
~\cite{Husain2023}.

Progress in EELS instrumentation has coincided with parallel progress in theoretical  understanding interacting electron systems. 
As we discuss below in Sections II-III, EELS measures the imaginary part of the dynamic charge susceptibility, $\chi''(q,\omega)$, which is the propagator for charge excitations in a many-body system~\cite{Platzman1973}. This quantity is fundamentally important in its own right, and played a critical role in benchmarking modern computational techniques, including local field corrections to the random phase approximation (RPA)
~\cite{Giuliani2005,Sturm1982}, 
which accounts for the heterogeneous property of real materials, and time-dependent density functional theory (TDDFT), which extended ground-state DFT techniques to excited states 
~\cite{Quong1993}. 
Knowledge of the dynamic charge response also played a critical role in the development of the GW approximation
~\cite{Reining2018,Onida2002}, 
which accounts for the dressing of individual quasiparticles by two-particle excitations, and was the first technique to accurately compute the band gap of simple semiconductors
~\cite{Louie1985,Louie1986}.

The purpose of this Review is to summarize the current state of EELS techniques, how they should be used, in general, to study valence band excitations and the influence of many-body interactions in the electron liquid of quantum materials, and what future investments in instrument development are needed to realize the full scientific potential of this technique. 

In Section II  of this manuscript, we start by discussing the essential response functions in condensed matter and different techniques for measuring them. In Section III we look in more depth at the charge response and how it reveals the screening properties and compressibility of a material. Sections IV and V review EELS techniques in transmission and reflection geometry, including the effect of surface losses that are always present in EELS measurements. Section VI summarizes classic EELS measurements on free-electron metals, such as Na and K, and Section VII summarizes the current state of affairs understanding strongly interacting and strange metals. We close, in Section VIII, by discussing what we feel are the most important directions for the future. 

\section{II. Response Functions for Condensed Matter}\label{sec:intro}

 Much of our understanding of condensed matter depends on knowledge of a few fundamental Green's functions that quantify how excitations propagate in a material \cite{Coleman2015,PaulMartin1968,Boothroyd2020}. The first is the retarded one-electron Green's function, 

\begin{equation}
    G(r,r',t-t') = -i \langle \{ \hat\psi^\dagger(r,t),\hat\psi(r',t') \}  \rangle \theta (t-t')/ \hbar,
\end{equation}

\noindent which represents the probability that an electron placed at location $r'$ at time $t'$ will propagate to location $r$ at some later time $t$. Here, $\hat\psi(r,t)$ is a Fermion annihilation operator, $ \{ , \} $ is an anticommutator, and $ \langle O \rangle = \sum_n  \bra{n}  \hat O \ket{n} P_n$ represents a quantum mechanical thermal average, $P_n = e^{-E_n/k_bT}$ being a Boltzmann factor. In principle, $G(r,r',t-t')$ is an independent function of $r$ and $r'$, but if the system has translational invariance, it will only depend of the difference, i.e., $r-r'$. 

The second is the density response, 

\begin{equation}
    \chi(r,r',t,t') = -i \langle [ \hat\rho(r,t),\hat\rho(r',t') ] \rangle \theta (t-t')/ \hbar,
\end{equation}

\noindent 
which represents the probability a disturbance in the charge density at $(r',t')$ will propagate to $(r,t)$. Here, $\hat{\rho}(r,t)$ is a bosonic operator for the charge density, and $[,]$ is a now a commutator. Finally, there is the magnetic response,

\begin{equation}
    \chi_{\sigma \sigma'}(r,r',t,t') = -i \langle [ \hat M_\sigma(r,t),\hat M_{\sigma'}(r',t') ] \rangle \theta (t-t')/ \hbar,
\end{equation}

\noindent which represents the propagation of disturbances in the local magnetic moment, $\hat{M}_\sigma$ being an operator for its $\sigma$ component.

The most informative and impactful probes of condensed matter measure one of these three Green's functions. For example, the DC resistivity of a material is determined by the $\omega=0$ and $q=0$ value of the charge response, eq. 2. A magnetometer, which measures the magnetic susceptibility, essentially measures the the $\omega=0$ and $q=0$ value of eq. 3. 

Two of the most important spectroscopic probes of materials, angle-resolved photoemission (ARPES) \cite{Damascelli2003} and scanning tunneling microscopy (STM), measure the one-electron Green's function (eq. 1). The ARPES intensity is proportional to the one-electron spectral function, $A(k,\omega) = - \operatorname{Im}{[G(k,k,\omega)]}/\pi$, where $G(k,k',\omega)$ is the Fourier transform of $G(r,r',t-t')$. Similarly, the tunneling probability in an STM experiment is proportional to $A(r,\omega) = - \operatorname{Im}{[G(r,r,\omega)]}/\pi$.
The tremendous impact these two probes have had in condensed matter physics is due to their measuring $G$ at the energy and momentum scales most relevant to quantum phenomena in materials, namely $\hbar \omega \ll k_B T$ (=25 meV at room temperature), and $k \ll 1/a$, where $a$ is the material lattice parameter. The impact of inelastic neutron scattering, which measures the magnetic response, eq. 3, with comparable resolution, has been similar and for the same reasons \cite{Boothroyd2020}. 

\section{III. Importance of the Charge Response}\label{charge}
The charge response, $\chi$, is perhaps unique in the number of different important physical phenomena it describes. In a homogeneous, 3D system, $\chi(r,r',t-t') = \chi(r-r',t-t')$ and it is customary to represent it in terms of its Fourier transform, $\chi(q,\omega)$. First and foremost, the fundmental collective charge modes in a many-body system, such as phonons, plasmons, excitons, etc., are defined as poles in the complex density response, $\chi(q,\omega)$, which manifest as peaks in its imaginary part. Further, the density response describes the ability of a material to screen charge, i.e., 

\begin{equation}
\epsilon(q,\omega) = \frac{1}{1+V(q) \chi(q,\omega)},
\end{equation}

\noindent where $\epsilon(q,\omega)$ is the longitudinal part of the dielectric function of the material, $V(q) = 4 \pi e^2/{q^2}$ being the 3D Coulomb interaction. Equivalently, $\chi(q,\omega)$ is related to the polarizability of the material,

\begin{equation}
\Pi(q,\omega) = \frac{\chi(q,\omega)}{1+V(q)\chi(q,\omega)},
\end{equation}

\noindent which quantifies the macroscopic polarization induced by an electric field, 

\begin{equation}
P(q,\omega) = - V(q) \Pi(q,\omega) E(q,\omega),
\end{equation}

\noindent the dielectric function being given by $\epsilon(q,\omega) = 1 - V(q) \Pi(q,\omega)$. 
Further, $\chi$ quantifies, in linear response, the charge density induced in a medium by an external charge, i.e., 

\begin{equation}
\rho_\mathrm{ind}(q,\omega) = V(q) \chi(q,\omega) \rho_\mathrm{ext}(q,\omega).
\end{equation}

Finally, the charge response defines, in principle, whether a material is a metal or an insulator. In the long wavelength limit, 

\begin{equation}
\lim_{q \rightarrow 0} \Pi(q,0) =  \frac{n^2}{V} \left ( \frac{\partial V}{\partial P} \right )_n = -n^2 \kappa,
\end{equation}

\noindent where $\kappa$ is the compressibility of the system, which vanishes for an insulator and is finite in a conductor. Eq. 8 is known as the compressibility sum rule \cite{Mahan2000}. Ultimately, the magnitude of the response is limited by the number of charges in the system, a fact quantified by the $f$-sum rule, 

\begin{equation}
\int_0^\infty \omega \operatorname{Im} \chi(q,\omega) d\omega = \frac{\pi n q^2}{2 m}.
\end{equation}

 Note that, because $\chi$ is a retarded Green's function, it satisfies the Kramers-Kronig relations, 

\begin{equation}\label{eq:kram1} 
\operatorname{Re} \chi(q,\omega) =\frac{2}{\pi} P \int\limits_0^{\infty} \frac{d\omega^\prime \omega' \operatorname{Im}\chi(q,\omega)}{(\omega^\prime)^2-\omega^2} 
\end{equation}

\begin{equation}\label{eq:kram1} 
\operatorname{Im} \chi(q,\omega) = - \frac{2 \omega}{\pi} P \int\limits_0^{\infty} \frac{d\omega^\prime \operatorname{Re}\chi(q,\omega)}{(\omega^\prime)^2-\omega^2},
\end{equation}

\noindent where we have used the fact that the charge density is real, i.e., $\chi^*(q,\omega) = \chi(q,-\omega)$ \cite{invnote}. 
Hence, sufficient knowledge of $\operatorname{Im}\chi$ would allow one to determine $\operatorname{Re}\chi$, and {\it vice versa}.

\section{IV. Inelastic Electron Scattering}

The greatest strength of electron scattering is that, under ideal conditions, it directly measures the dynamic charge response, eq. 2. The ultimate aim for modern EELS techniques is to do so with energy and momentum resolution comparable to ARPES and INS. To understand to what extent this is possible, we will start with an idealized picture of inelastic electron scattering and then discuss to what extent it applies in different experimental implementations. Assuming the first Born approximation holds, the matrix element for scattering of electrons, in the interaction picture, is given by 

\begin{equation}
M = - \frac{i}{\hbar}  \bra{f} H'(0) \ket{i}.
\end{equation}

\noindent Here, the initial and final states $\ket{i} = c_{k_i}^\dagger \ket{m}$ and $\ket{f} = c_{k_f}^\dagger \ket{n}$, where $\ket{m}$ and $\ket{n}$ are many-body eigenstates of the material system, including both electron and ion degrees of freedom. $k_i$ and $k_f$ are momenta of the probe electron, which is assumed to always reside in a plane wave state. The relevant interaction is Coulomb, 

\begin{equation}
H'(0) = \frac{1}{2} \int{\frac{\hat\rho(r_1) \hat\rho(r_2)}{|r_1-r_2|} dr_1^3 dr_2^3}
\end{equation}

\noindent where $\hat \rho(r)$ is an operator for the total charge density, of both electrons and ions, at location $r$. 
Neglecting exchange scattering and evaluating $M$ gives

\begin{equation}
M_{n,m} = \frac{e}{2 i \hbar v} \int{ \frac{\bra{n} \hat \rho(r_1) \ket{m} e^{i(k_i-k_f)\cdot r_2}}{|r_1-r_2|} dr_1^3 dr_2^3 }
\end{equation}

\noindent where $e$ is the charge of the probe electron and $v$ is the volume of all space. Shifting its center, $r_2 \rightarrow r_2+r_1$, and doing the $r_2$ integral gives



\begin{equation}
M_{n,m} = \frac{e}{2 i \hbar v} \bra{n} \hat\rho(q) \ket{m} \frac{4 \pi}{q^2}
\end{equation}

\noindent where we have identified $q = k_f - k_i$ as the momentum transferred to the probe electron. The quantity $\bra{n} \hat\rho(q) \ket{m}$ represents the many-body transition charge density of the electron+ion system. 

To turn this into a scattering cross section at finite temperature, we apply Fermi's golden rule, 

\begin{equation}
\frac{\partial^2 \sigma}{\partial \Omega \partial E} = \frac{1}{\Phi} \sum_{n,m} w_{n \leftarrow m} P_m \frac{\partial^2 N}{\partial \Omega \partial E}
\end{equation}

\noindent where $\Phi = \sqrt{2E_i/m} / v$ is the incident electron flux, $E_i$ being the incident electron kinetic energy, and $\partial^2 N / \partial \Omega \partial E$ is the density of final states. The transition rate is given by 

\begin{equation}
w_{n \leftarrow m} = 2 \pi \hbar |M_{n,m}|^2.
\end{equation}

\noindent The density of final states is given by the usual expression,

\begin{equation}
\frac{\partial^2 N}{\partial \Omega \partial E} = \frac{v}{8\pi^3} \left ( \frac{2m}{\hbar^2} \right ) ^{3/2} \sqrt{E} .
\end{equation}

\noindent Multiplying out all the terms, putting in the energy-conserving delta functions, gives 

\begin{equation}
\frac{\partial^2 \sigma}{\partial \Omega \partial E} = \sqrt{\frac{E_f}{E_i}} \frac{2 m^2 e^2}{\hbar^4 q^4} S(q,\omega)
\end{equation}

\noindent where 

\begin{equation}
S(q,\omega) = \sum_{n,m} { | \bra{n} \hat \rho(q) \ket{m} |^2 P_m } \delta (\hbar \omega-E_n+E_m) 
\end{equation}

\noindent is the dynamic structure factor. Also called the Van Hove function \cite{vanHove1954}, $S(q,\omega)$ is the Fourier transform of the density autocorrelation function, 

\begin{equation}
S(r,t) = \int{\langle  \hat\rho(r',t') \hat\rho(r'+r,t'+t) \rangle}  dr'^3 dt'.
\end{equation}

It is here that we see the true value of EELS as a probe of condensed matter. The Van Hove function is directly related to the density Green's function by the quantum mechanical version of the fluctuation-dissipation theorem, 

\begin{equation}
S(q,\omega) = - \frac{\hbar}{\pi} \frac{1}{1-e^{-\hbar \omega/k_B T} } \operatorname{Im}{\chi(q,\omega)},
\end{equation}

\noindent the prefactor $(1-e^{\hbar \omega/k_B T})^{-1}$ being the Bose factor.
EELS therefore directly measures the density Greens function, revealing the collective charge excitations of the system, its finite-$q$ screening properties, and quantifying its polarizability and compressibility. Note that 
$\operatorname{Im}\chi(q,\omega) < 0$ at $\omega>0$, so the correlation function, $S(q,\omega)$ is a positive-definite quantity.  

Expressed in terms of the dielectric function (using eq. 4), eq. 16 has the form

\begin{equation}
\frac{\partial^2 \sigma}{\partial \Omega \partial E} = \sqrt{\frac{E_f}{E_i}} \frac{m^2}{2\pi^2 \hbar^3 q^2} \frac{1}{1-e^{-\hbar \omega/k_B T} }
\operatorname{Im}{\frac{-1}{\epsilon(q,\omega)}},
\end{equation}

\noindent where the quantity $- \operatorname{Im}[1/\epsilon(q,\omega)]$ is called the ``loss function." Eq. 20 shows, generally, that the EELS cross section decreases like $1/q^2$, which is intrinsic to any measurement in which the coupling is mediated by Coulomb. Further, because the Bose factor is singular at $\omega \sim 0$, EELS measurements tend to show a large peak at $\omega=0$. Sometimes called the ``elastic line," this spectral feature is part of the correlation function and an essential component of the material response. 

\section{V. EELS Techniques}

There are two basic configurations for performing EELS experiments, transmission and reflection (Fig. 1). The former is typically done at high energy, 30-300 keV, as transmission through suspended thin films. The latter is normally done at low energy on ultraclean surfaces. Both approaches are experimentally challenging and have different advantages and disadvantages. 

\begin{figure}
    \centering
    \includegraphics[width=0.45\textwidth]{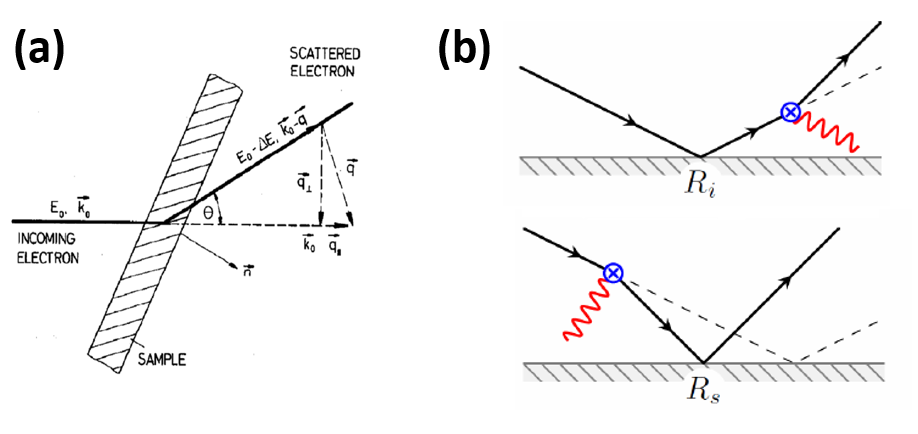}
    \caption{(a) Geometry of an inelastic electron scattering experiment in transmission through a thin film (T-EELS). (b) Geometry of an inelastic electron scattering experiment in reflection (R-EELS). The latter is mediated by a specular reflection that supplies the momentum perpendicular to the surface \cite{Mills1975,Ibach1982,Kogar2014,Vig2017}
    }
    \label{Fig1}
\end{figure}

\subsection{Transmission Geometry}

Transmission EELS (T-EELS) experiments have been performed using both dedicated spectrometers \cite{Boersch1964,Schroder1972,Raether1965,Schnatterly1979,Fink1989} and with transmission electron microscopes \cite{Egerton1996,Egerton2008,Colliex2022}. The general configuration of a T-EELS measurement is shown in Fig. 1, with the electron momentum vectors, $k_i$ and $k_f$ indicated. Following the scattering event, the energy and momentum of the scattered electron is measured with an energy analyzer. These experiments need to be done at sufficiently high energy that transmission through a suspended film is possible. For a beam energy 30-300 keV, the film thickness should not exceed $\sim$100 nm.

The biggest advantage of T-EELS is that, because the electrons pass through the bulk of the material, at least part of the cross section resembles Section IV and is proportional to the bulk density response, $\operatorname{Im}\chi(q,\omega)$. That said, because the materials are so thin, excitations from the surface are known to make a significant contribution to the spectrum, though well-posed strategies for separating bulk from surface losses by measuring samples with varying thickness have been demonstrated \cite{mkhoyan2007}. 

For both dedicated and STEM-based spectrometers, the energy resolution was usually limited by the spectral width of the electron source, yielding total energy resolutions $\Delta E \approx$ 0.1 eV, though, notably, 4 meV resolution was achieved in one early study using Wien filters \cite{Schroder1972}. More recently, using aberration correction techniques originally developed for achieving sub-$\AA$ spatial resolution, energy resolution better than 10 meV has been achieved in a STEM setup~\cite{Krivanek2014}. 

At the energy scale of a few meV, excitations tend to exhibit a high degree of quantum nonlocality \cite{Husain2023}, and it becomes critical to perform EELS measurements with high momentum resolution rather than high spatial resolution, the tradeoff being limited by Liouville's theorem. Recognizing this, early, dedicated spectrometers were built to achieve high momentum resolution, typically  $\Delta q \sim 0.04 \AA^{-1}$. This was not easily done; at such high energies, the primary electron momentum $k_i \sim 60-200 \AA^{-1}$, meaning an angular resolution of a few milliradians must be achieved to reach momentum transfers relevant to a physical material, which should be a small fraction of a Brillouin zone ($\sim \AA^{-1}$). Achieving high momentum resolution necessarily requires poor spatial resolution, and is intrinsically incompatible with high resolution imaging. 

By contrast, modern, meV-resolved STEM instruments, which are still generally configured to achieve sub-$\AA$ focusing, have extremely poor momentum resolution, usually exceeding a full Brillouin zone of a typical material \cite{Husain2019,Husain2023}. This, unfortunately, has the effect of washing out the very excitations such excellent energy resolution is designed to observe. Transmission EELS measurements of charge collective modes appears, at the moment, to be a lost art.

\subsection{Reflection Geometry}

Reflection EELS (R-EELS) measurements are mainly performed with much lower beam energy $E_i\sim$ 10-100 eV, at which the probe electron never actually enters the material, and instead reflects off the surface \cite{Ibach1982}, also illustrated in Fig. 1. In R-EELS measurements the sample thickness becomes unimportant; what is critical is the surface quality, which must be pristine at nearly the atomic level, requiring {\it in situ} cleaving or other surface preparation techniques. 

The biggest advantage of R-EELS is that, because the beam energy is so low, it is possible to achieve very high energy resolution; better than 1 meV has been achieved \cite{Ibach1993}, and 5 meV is typical \cite{Ibach1991}. Further, it is quite straight forward to achieve the relevant momentum resolution, $\Delta q \sim 0.01 \AA^{-1}$, which is a small fraction of a typical Brillouin zone \cite{Husain2023}. 

The subtlety of R-EELS is that it measures a different charge response than that described in Section IV \cite{Kogar2014,Vig2017,Ibach1982}, 

\begin{equation}
\operatorname{Im}{\chi''_{\mathrm{s}}(q,\omega)} = \int_{-\infty}^0 dz_1 \int_{-\infty}^0 dz_2 \, e^{-q|z_1+z_2|} \operatorname{Im}{\chi(q,\omega;z_1,z_2)},
\end{equation}

\noindent where $\chi(q,\omega;z_1,z_2)$ is a mixed version of eq. 2 in which $q$ now represents the momentum parallel to the surface \cite{Kogar2014,Vig2017,Ibach1982}. Sometimes called the ``surface response,'' eq. 24 represents the bulk response of a semi-infinite system as measured through its surface, $z_1$ and $z_2$ representing the depth into the material, emphasizing that the momentum perpendicular to the surface, $q_z$ is no longer conserved. Like T-EELS, R-EELS contains a mixture of bulk and surface excitations, the relative weight of which depends upon the magnitude of $q$ \cite{Chiarello2000,Kogar2017,Husain2023}. Note, however, that eq. 24 does not reduce to eq. 23 in any limit, not even as $q \rightarrow 0$. Nevertheless, if $\chi_s$ is assumed to be independent of $q$, the $z$ integrals may be done explicitly and $\operatorname{Im}{\chi''_{\mathrm{s}}(q,\omega)} \propto -\operatorname{Im}{1/[1+\epsilon(\omega)]}$, which is usually called the surface loss function, $\epsilon(\omega)$ representing the {\it bulk} dielectric function \cite{HusainThesis2020,Nazarov1994}. Eq. 24 may therefore be related to the bulk response through material-specific modeling \cite{Chen2024,Ibach1982}.

The original implementations of R-EELS used single-point spectrometers based on electrostatic deflectors with cylindrical \cite{Ibach1991} and, later, toroidal geometry \cite{Ibach1993}. Similar to a triple-axis neutron spectrometer, these setups measure a single $q$ and $\omega$ at a time, spectra being generated by stepping the lenses and scattering angle \cite{Ibach1991}. The momentum resolution and accuracy of these setups was later improved by employing eucentric goniometers and improved tuning techniques, an approach sometimes called M-EELS \cite{Vig2017,Kogar2017,HusainThesis2020,Husain2023,Chen2024}. Much faster data acquisitions was demonstrated by employing an ARPES-style hemispherical analyzer, which provides parallel readout of both energy and momentum \cite{Ibach2017,Li2023,Zhu2015}. Even greater speeds could be achieved by employing time-of-flight techniques. 
There is therefore vast room for straight-forward improvement of R-EELS techniques using currently available technologies. 

\section{VI. Nearly-free electron metals}\label{section:IV}

The characteristic collective mode of a metal is the plasmon, which is a longitudinal excitation of the electron density analogous to a sound wave. In neutral systems, such as an atomic gas, longitudinal modes are gapless or acoustic. Plasmons in charged systems, however, are gapped by the long-ranged Coulomb interaction, the transverse plasmons remaining gapless. This generation of mass by interactions, pointed out by Schwinger \cite{Schwinger1962} and Anderson \cite{Anderson1963}, helped inspire the Higgs mechanism in particle physics \cite{Higgs1964}. 

\subsection{Homogeneous Materials}

The textbook method for describing plasmons is the Random Phase Approximation (RPA)  \cite{Pines1952,Pines1966}, which treats a plasmon as electron-hole pairs interacting through direct Coulomb. Plasmons appear as zeros in the dielectric function, $\epsilon(q,\omega)=0$, which in RPA gives a plasma frequency $\omega_p^2= 4 \pi n e^2/m^*$, $n$ being the density and $m^*$ the effective mass. A plasmon in RPA has infinite lifetime and disperses like $q^2$ until it hits the single-particle continuum at the critical momentum, $q_c$, above which it decays by Landau damping \cite{Pines1966}. 

Early EELS experiments on simple metals generally supported the RPA picture. Consistent with expectations from Section V, T-EELS on Al, K, and Na, showed both bulk and surface plasmons \cite{Batson1983,Felde1989}, the latter appearing at the Ritchie frequency, $\omega_R \sim \omega_p / \sqrt{2}$ \cite{Ritchie1957,Plummer1995}. It was found, generally, that the surface plasmon faded with increasing momentum, the bulk mode dominating the spectra at larger $q$, though it is not established if this effect applies to phonons, excitons, or other types of excitations.
The dispersion of the bulk mode is quadratic, as expected, though with a lower coefficient ($\alpha=0.17$ for Na compared to the RPA value of 0.32). Further, the width undergoes an abrupt increase at roughly the correct value of $q_c$ \cite{Felde1989,Batson1983}. 

The situation with R-EELS is similar. Both bulk and surface modes are visible, with the bulk mode becoming more pronounced at larger $q$ \cite{Chiarello2000}. Experiments on simple, doped semiconductors find that the surface mode appears close to the Ritchie value \cite{Kogar2015}. A peculiarity of the surface mode is that its dispersion has a term that is linear $q$ that can be positive or negative \cite{Tsuei1989,Sprunger1992}. This was eventually explained as a consequence of the location of the induced charge with respect to the surface \cite{Feibelman1982,Silkin2004}. 

A major early concern was the plasmon linewidth in the long-wavelength limit, which in RPA is zero, but in the experiment is finite. Including beyond-RPA corrections did not seem to address this discrepancy \cite{DuBois_1969}. In simple alkali halides, Na and K, the width was eventually explained as arising from decay into interband transitions assisted by Umklapp scattering \cite{Paasch1970,Gibbons1977,Felde1989,Eguiluz1999}. But in correlated metals this width is still a major outstanding issue, as discussed further below. 

\subsection{Layered Materials}

A significant number of materials of interest today are layered and quasi-two dimensional. Collective modes in layered materials have some important, idiosyncratic features, which we briefly review here.

The long-ranged Coulomb interaction is given by $V(r)=e^2/r$, independent of the dimensionality or geometry of a material. In a homogeneous, 3D system, this enters as the Fourier transform, $V(q) = 4\pi e^2/q^2$ (Sections II-III above). In 2D materials, what is relevant is the mixed transform, $V_{2D}(q,z)=(2 \pi e^2/q) e^{-q|z|}$, where $q$ is the in-plane momentum and $z$ the height above the layer. In a periodic, layered material, the form that enters \cite{JainAllen1985} is

\begin{equation}
V_{\mathrm{L}}(q,q_z)=\frac{2\pi e^2}{q} \frac{\sinh{qd}}{\cosh{qd}-\cos{q_z d}},
\end{equation}

\noindent where $d$ is the layer spacing, $q$ is the in-plane momentum, and $q_z$ is a periodic variable representing the crystal momentum perpendicular to the layers. 

For a noninteracting electron gas, the polarizability $\Pi = \omega_p^2 q^2 d / 4 \pi e^2 \omega^2$ \cite{JainAllen1985}, in which case the dielectric function $\epsilon = 1 - V_{\mathrm{L}} \Pi \rightarrow 1-\omega_p^2/\omega^2$ in the long wavelength limit. Setting $\epsilon(q,\omega)=0$ gives the full momentum dependence of the plasma frequency \cite{JainAllen1985},

\begin{equation}
\omega_p(q,q_z) = \left [ \omega_p^2 \frac{qd}{2} \frac{\sinh{qd}}{\cosh{qd}-\cos{q_z d}} \right ]^{1/2}.
\end{equation}

\noindent Originally derived by Fetter \cite{Fetter1974}, eq. 26 leads to the remarkable conclusion that, while $\omega_p$ is nonzero in the long wavelength limit, it vanishes as a function of $q$ for any nonzero value of $q_z$. Eq. 26 is sometimes discussed, colloquially, as having acoustic and optic branches. However, the expression has only a single branch, but exhibits nonanalytic behavior, with $q \rightarrow 0$ and $q_z \rightarrow 0$ being noncommuting limits (ultimately, the finite size of the sample will round out this singular behavior). Eq. 26 is usually discussed in the context of plasmons, but the effect is generic and applies to any kind of longitudinal collective mode in a layered system. 

\section{VII. Highly Correlated Metals}

In lower density metals, such as heavier alkalis with larger lattice constants, interactions are more important and subtleties arise. In Cs, for example, the bulk plasmon dispersion is negative \cite{Felde1989}, defying expectations from RPA. This effect was initially thought to arise from an incipient Wigner crystal, but was eventually explained by including vertex corrections not captured by RPA \cite{Aryasetiawan1994,Eguiluz1997}. 

\subsection{Strange Metals}

For the case of the strange metals, the challenges become severe. Suitable, beyond-RPA descriptions of strange metals do not yet exist, and T-EELS and R-EELS experiments give contradictory results at nonzero values of $q$ (Fig. 2).

\begin{figure}
    \centering
    \includegraphics[width=0.45\textwidth]{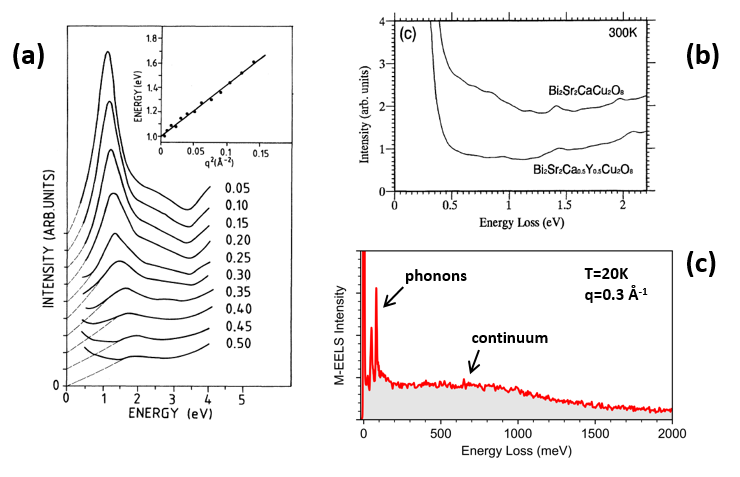}
    \caption{Three different EELS measurements of Bi-2212 yielding contradictory results. (a) T-EELS measurements of N{\"u}cker \cite{Nuecker1989} at 170 keV showing a dispersing, RPA-like plasmon characteristic of a weakly interacting electron gas. (b) T-EELS measurement from Terauchi \cite{Terauchi1995,Terauchi1999} at 60 keV at $q \sim 0.3 \AA^{-1}$ and resolution 39 meV showing a continuum instead of a plasmon. (c) R-EELS measurement at 50 eV, also at $q = 0.3 \AA^{-1}$, that also shows a continuum, albeit with a slightly different shape \cite{Mitrano2018,Husain2019,Chen2024}. }
    \label{Fig1}
\end{figure}

The strange metal was first observed in the normal state of copper-oxide high temperature superconductors, and since has been observed in many correlated metals including Bechgaard salts \cite{Doiron2009}, $\beta$-YbAlB$_4$ \cite{Tomita2015}, Ge-doped YbRh$_2$Si$_2$ \cite{Trovarelli1999}, bilayer Sr$_3$Ru$_2$O$_7$ \cite{Mousatov2020}, CeCoIn$_5$ \cite{Tanatar2007}, Ba(Fe$_{1-x}$Co$_x$)$_2$As$_2$ \cite{Doiron2009}, and Magic-angle bilayer graphene \cite{Cao2020}, among others. Its main experimental signature is a resistivity that is linear in $T$ and does not saturate at the Mott-Ioffe-Regel (MIR) limit \cite{Hussey2004}. This implies a scattering rate, $\tau^{-1} = k_B T/\hbar$, limited only by fundamental constants and not by specifics of the material itself. Now referred to as ``Planckian dissipation'' \cite{Zaanen2019}, this scattering rate is thought to represent a fundamental bound analogous to the cosmological Planck time \cite{Bruin2013}. Other defining attributes of strange metals are a quasiparticle lifetime $\Gamma \sim \sqrt{(\hbar \omega)^2 + (k_B T)^2}$ \cite{Reber2019} and a frequency-dependent conductivity with a power law form \cite{vdMarel2003}. 

These properties inspired the marginal Fermi liquid (MFL) hypothesis, which proposed that the electrons were coupled to a continuum of charge excitations of unknown origin, whose response is constant in frequency and independent of momentum \cite{Varma1989}. MFL was able to account for all the above properties, but was utterly at odds with RPA, which predicts a dispersing plasmon with $\omega_p \sim 1$ eV \cite{Mitrano2018}. 

Infrared optical experiments, which measure the dielectric function at $q=0$, do indeed show a clear plasmon with frequency close to 1 eV \cite{vdMarel2003,Levallois2016}. Further, RIXS experiments at nonzero $q_z$ show evidence for a low-frequency plasmon consistent with the Fetter model, eq. 26 \cite{Hepting2018,Nag2020,Hepting2022}.

However, the simple RPA picture does not hold up upon close scrutiny. The plasmon has a broad lineshape at $q=0$, consistent with a power law conductivity $\sigma \sim \omega^{-2/3}$ and renormalized scattering rate with an MFL form, $(\tau^*)^{-1} \sim \omega$ \cite{vdMarel2003}. Further, the plasma frequency does not shift with density, as it should for a free-electron metal \cite{Levallois2016}. These peculiarities emphasize the urgency measuring the density response at nonzero momentum, where the differences between RPA and MFL are most pronounced. 

Early T-EELS efforts to measure the momentum dependence of the plasmon in Bi$_2$Sr$_2$CaCu$_2$O$_{8_x}$ (Bi-2212), a strange metal, led to conflicting results (Fig. 2). N{\"u}cker and coworkers observed a plasmon that is in perfect agreement with optics at $q=0.05 \AA^{-1}$, but exhibited completely ordinary, quadratic RPA-like dispersion (Fig. 2), clearly contradicting MFL \cite{Nuecker1989}. This result was reproduced shortly thereafter by Wang \cite{Wang1990}. However, subsequent studies by Terauchi at a momentum $q\sim 0.3\AA^{-1}$ showed no plasmon at all, but rather a frequency-independent continuum reminiscent of the MFL picture \cite{Terauchi1995,Terauchi1999}.

Later, R-EELS was done on the same material by Schulte, who observed a 1 eV plasmon in the small momentum regime, though with a somewhat different lineshape from T-EELS or optics \cite{Schulte2002}. Eventually Chen and coworkers \cite{Chen2024} reproduced these results with higher resolution and, using the layered model of Jain and Allen \cite{JainAllen1985}, showed them to be in quantitative agreement with IR optics in the optical limit, $q<0.04 \AA^{-1}$. Focusing on the large $q$ regime, Mitrano observed a frequency- and momentum-independent continuum, with a cutoff around 1 eV, very similar to ref. \cite{Terauchi1999} and reminiscent of the MFL hypothesis \cite{Varma1989}. Performing a full composition dependence, Husain showed that this continuum correlates closely with strange metal behavior in the Bi-2212 phase diagram \cite{Husain2019}. 

The overall picture that emerges is one in which all experimental probes---T-EELS, R-EELS, and IR optics---are quantitatively consistent at $q \sim 0$. But different EELS measurements give profoundly different and contradictory results at larger $q$. Theoretical understanding of density fluctuations in strange metals will not be possible until these experimental discrepancies are resolved. 

\subsection{The intermediate case of Sr$_2$RuO$_4$}

An enlightening comparison can be made by considering the multiband metal Sr$_2$RuO$_4$, which is intermediate between a normal and a strange metal. At low temperature, $T \lesssim 40$K, Sr$_2$RuO$_4$ is an excellent Fermi Fermi liquid showing resistivity $\rho \sim T^2$, well-defined quantum oscillations \cite{Mackenzie2003}, and Fermi liquid scattering rate $\tau^{-1} \sim (\hbar \omega)^2 + (k_B T)^2$ in optics \cite{Stricker2014}. However at higher temperatures, $T \gtrsim 600$ K, Sr$_2$RuO$_4$ crosses over into a strongly interacting phase in which the quasiparticles are highly damped \cite{Wang2004}, the resistivity $\rho \sim T$, and the MIR limit is violated at high temperature \cite{Tyler1998}. The strong interactions arise from Hund’s coupling and are described well by dynamical mean field theory \cite{Tamai2019}. One therefore expects the excitations in Sr$_2$RuO$_4$ to be Fermi liquid-like at low energy, and to cross over to something more strange metal-like at energy scales greater than $\sim$0.1 eV. 

Again, consistency between T-EELS and R-EELS measurements for this material is poor. Low-energy R-EELS measurements do indeed observe a crossover between Fermi liquid behavior and more correlated behavior as a function of energy. At energy scales below 0.1 eV, the primary excitation is an out-of-phase plasmon, known as a ``demon" \cite{Pines1956}, which has an acoustic dispersion in reasonable agreement with RPA \cite{Husain2023}. At higher energy scales, and larger $q$, R-EELS measurements show a MFL-like continuum with a cutoff of $\sim$1.5 eV, consistent with strange metal behavior at high energy. 

By contrast, T-EELS measurements show Sr$_2$RuO$_4$ to be a weakly interacting Fermi liquid at all energy scales, with a completely ordinary, RPA-like plasmon with a plasma frequency $\omega_p = 1.5$ eV \cite{Knupfer2022}. For $q_z=0$, this plasmon exhibits textbook, $q^2$ dispersion, and a critical momentum of about 0.5 $\AA^{-1}$. For $q_z \neq 0$, its behavior adheres well to the Fetter model (eq. 26), the plasma frequency energy falling rapidly toward zero at $q \rightarrow 0$ \cite{Schultz2024}. Some of this behavior is quite surprising since the plasmon does not exhibit an RPA lineshape even in IR optics, which show extra spectral weight above 0.1 eV requiring DMFT to explain \cite{Stricker2014}. 

\section{VIII. Summary and Outlook}

The momentum-dependent two-particle charge response, $\chi(q,\omega)$, is critical to our understanding interacting electron materials, in one, two, and three dimensions, providing information about collective modes, screening properties, and the compressibility. We have reviewed efforts over the last decades to measure this quantity in a selection of materials by means of inelastic electron scattering in both transmission and reflection geometry. In simple metals, such as Na, K, and Al, the random phase approximation provides a reasonable understanding of the charge response. But in more correlated metals, such as Cs, corrections are required. 

In the case of strongly interacting and strange metals, different experiments currently yield different results, with some EELS measurements reporting a simple, RPA-like plasmon, and others reporting a continuum reminiscent of the marginal Fermi liquid hypothesis of the late 1980's. The biggest discrepancies occur at larger momenta, where experiments are more challenging. Future understanding of strongly interacting metals will not be possible without improved measurements that resolve these discrepancies. 

\begin{figure}
    \centering
    \includegraphics[width=0.45\textwidth]{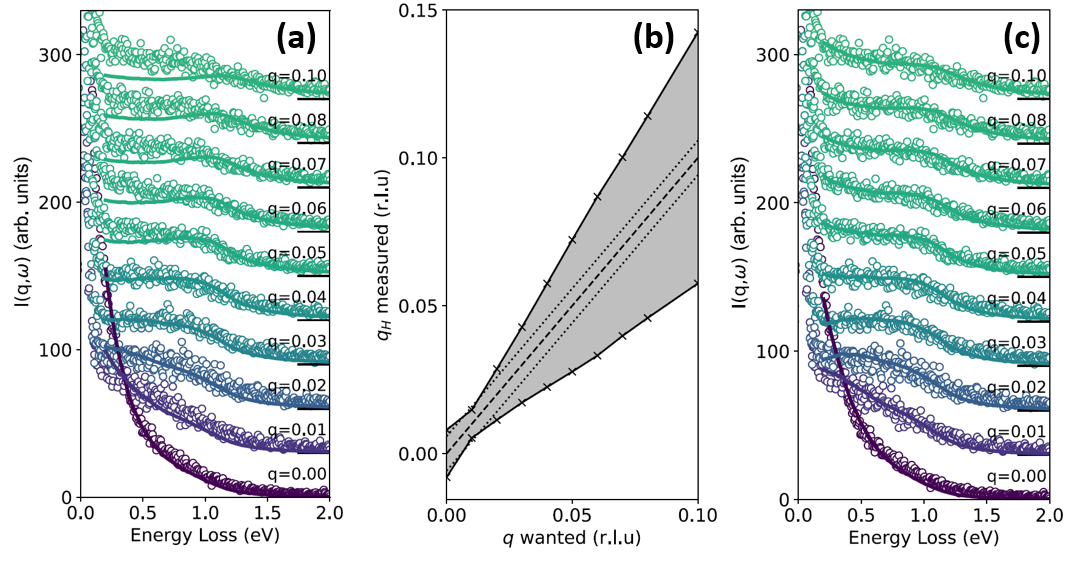}
    \caption{Comparison between IR and R-EELS measurements of Bi-2212 in the low momentum regime, made by using the Jain-Allen formalism \cite{JainAllen1985}. The data are reproduced from ref. \cite{Chen2024}. (a) R-EELS data (open circles) compared to the surface response calculated from the IR measurements of ref. \cite{Levallois2016}, showing good agreement in the small $q$ limit. (b) Range of momenta used to ``scramble" the probe electron in the simulations. (c) Same comparison as (a) but assuming the momentum of the probe electron is, for some reason, not conserved, showing reasonable agreement across the whole momentum range. 
}
    \label{Fig1}
\end{figure}

An intriguing possibility is that the results depend on the energy scale of the measurement. EELS measurements at 170 keV (Fig. 1(a)) show an RPA plasmon, while similar measurements at 60 keV and 50 eV see a continuum (Fig. 2(b),(c)). In ref. \cite{Chen2024}, Chen and coworkers used the layered formalism of Jain and Allen \cite{JainAllen1985} to check the consistency between IR optics and R-EELS measurements in the long-wavelength limit. The strategy was to use the $\Pi(q,\omega)$ determined from IR experiments to calculate the surface response (eq. 24) and compare to R-EELS measurements. Explicit comparison resulted in Fig. 3(a), which shows good agreement for small momenta, $q < 0.07 \AA^{-1}$ (0.04 r.l.u.). At larger momenta, however, the results no longer agree, which is perhaps unsurprising since IR gives the value of the polarizability only at $q=0$.  

Strikingly, however, the authors found that if they artificially ``scrambled" the momentum of the measurement, as illustrated in Fig. 3(b), excellent agreement was obtained even at larger momenta (Fig. 3(c)). This suggests the intriguing possibility that all measurements may be correct, but that somehow translational symmetry is dynamically broken in strange metals in a way that particularly affects low-energy electrons. 

Improved measurements, with improved instruments, are greatly needed. The most important developments will be in the continued development of meV-resolved hemispherical and time-of-flight techniques for R-EELS measurements, and meV-resolved STEM-based T-EELS instruments with high momentum resolution. The latter will require researchers to sacrifice high spatial resolution, which will necessitate something of a culture shift among electron microscopists. 

\section{IX. Acknowledgements} 

\begin{acknowledgements}
We thank P. W. Phillips and A. Howie for helpful discussion and correspondence, A. Husain for preparing Fig. 2(c), and N. de Vries and X. Guo for proofing the manuscript. This work was supported by the EPiQS program of the Gordon and Betty Moore Foundation, grant GBMF9452.
\end{acknowledgements}

\bibliographystyle{apsrev4-2}
\bibliography{References}

\begin{thebibliography}{129}
\bibitem [{inv()}]{invnote}%
  \BibitemOpen
  \href@noop {} {}\bibinfo {note} {This assumes the system is inversion
  symmetric. The more general expression is $\chi^*(q,\omega) =
  \chi(-q,-\omega)$.}\BibitemShut {Stop}%
\end{thebibliography}

\end{document}